\documentclass[journal]{IEEEtran}

\ifCLASSINFOpdf
\else
   \usepackage[dvips]{graphicx}
\fi
\usepackage{url}

\usepackage{graphicx}
\usepackage{amsmath,amssymb,amsfonts}
\usepackage{siunitx}
\usepackage[caption=false,font=normalsize,labelfont=sf,textfont=sf]{subfig}

\begin{document}

\title{Uncertainty Propagation in Finite Impulse Response Filters: Evaluating the Gaussian Assumption}

\author{Jennie Couchman and Phillip Stanley-Marbell}

\maketitle

\begin{abstract}
A common assumption in signal processing is that underlying data numerically conforms to a Gaussian distribution.
It is commonly utilized in signal processing to describe unknown additive noise in a system and is often justified by citing the central limit theorem for sums of random variables, although the central limit theorem applies only to sums of independent identically distributed random variables.
However, many linear operations in signal processing take the form of weighted sums, which transforms the random variables such that their distributions are no longer identical.
One such operation is a finite impulse response (FIR) filter.
FIR filters are commonly used in signal processing applications as a pre-processing step.
FIR output noise is generally assumed to be Gaussian.
This article examines the FIR output response in the presence of uniformly distributed quantization noise.
We express the FIR output uncertainty in terms of the input quantization uncertainty and filter coefficients.
We show that the output uncertainty cannot be assumed to be Gaussian, but depending on the application a Gaussian estimation may still be useful.
Then, we show through detailed numerical simulations that the output uncertainty distribution of the filter can be estimated through its most dominant coefficients.

\end{abstract}


\section{Introduction}
\IEEEPARstart{I}{t} is a common assumption in signal processing paradigms that underlying data numerically conforms to a Gaussian distribution.
This assumption is often justified by referring to the central limit theorem (CLT) for sums of random variables. 
Let $\lbrace X_1, X_2, \ldots,X_N\rbrace$ be a sequence of $N$ independent and identically distributed (i.i.d.) random variables with mean $\mu$ and variance $\sigma^2$.
The CLT states that for large enough $N$ the distribution of the arithmetic mean of this sequence, 
\begin{equation}
\bar{X} = \frac{X_1 + X_2 + \ldots + X_N}{N},
\end{equation}
will tend towards a Gaussian distribution with mean $\mu$ and variance $ \sigma^2/N$.

Digital filters are ubiquitous in signal processing applications, generally as a pre-processing step. 
They are often used to examine specific frequency components of a given signal.
FIR filters are a subclass of digital filters whose impulse response has a finite duration.
The output $y\left[n\right]$ of an $N^{\text{th}}$-order FIR filter is
\begin{equation}
\label{eq:FIR_Resp}
y\left[n\right] = \sum_{i=0}^N b_i x\left[n-i\right].
\end{equation}
where $x\left[n\right]$ is a finite-length discrete-time input signal.
The filter coefficients $\left\lbrace b_i \right\rbrace$ completely characterize the filter's input response.
Equation~\ref{eq:FIR_Resp} defines the output of an FIR filter as the digital convolution of the filter coefficients and the input signal.

Even in the absence of noise in the input samples, digital signals contain at least some amount of uncertainty due to the process of quantizing the signal after it is sampled.
Quantization uncertainty is usually modeled as uniform random variables.
Equation~\ref{eq:FIR_Resp} then becomes a weighted sum of independent uniform random variables.
The outputs of an FIR filter are often assumed to conform to a Gaussian distribution, as researchers and practicing engineers often interpret the CLT as implying \emph{any} summation of a large number of random variables will tend towards a Gaussian distribution.
However, the CLT is defined specifically for i.i.d. variables. 
The output of an FIR filter is a \emph{weighted} sum of its input terms, so while the input variables can be considered independent, they can no longer be considered \emph{identically distributed}. 
There has been previous work in determining the conditions where the CLT can be applied to weighted sums~\cite{Kamgar-Parsi1995, Cuzick1995, Weber2006, Avena2024} but to our knowledge these analyses have not yet been investigated for their relationship to FIR filters.

In this paper we make the following contributions:

\begin{enumerate}
	\item Mathematically express the output uncertainty of an FIR filter in terms of its input uncertainty.
	\item Demonstrate that the output of an FIR filter cannot be assumed to be Gaussian.
	\item Demonstrate that the output uncertainty distribution of an FIR filter can be estimated by the most dominant filter coefficients.
\end{enumerate}

\section{FIR Filter Response to Quantization Noise}

In practice, the inputs to filters are measurements of some physical process and all measurements by definition have some associated measurement uncertainty in sampling the underlying measurand, as well as quantization uncertainty in converting that sampled measurand (the measurement) into a digital representation. As a result, it is natural to investigate how these measurement and quantization uncertainties are transformed by the arithmetic performed by the filter (Equation~\ref{eq:FIR_Resp}), to yield an uncertainty in the filter response.

\subsection{Quantization Uncertainty Model}
\label{sec:MeasurementUncertainty}
The input of the filter $x\left[n\right]$ is now viewed as a series of uncertain measurements. 
In particular, the $i$-th sample of $x\left[n\right]$ is modeled as a uniformly distributed random variable $X_i \sim \mathcal{U}\left(\mu_i-\frac{\delta}{2},\mu_i+\frac{\delta}{2}\right)$.
The probability density function (pdf) of $X_i$ is then
\begin{equation}
f_{X_i}\left(x_i\right) = 
	\begin{cases}
		\frac{1}{\delta} & \text{if } \lvert x_i - \mu_i \rvert < \frac{\delta}{2} \\
		0 & \text{otherwise}
	\end{cases}.
\end{equation}
The mean value of $X_i$ is $\mu_i$ and it is the quantization of the measured value and
$\delta$ is the quantization step size.
The underlying measurand is understood to take any value within the support of the random variable.
The series of measurements are all assumed to be mutually independent of each other.

\subsection{Uncertainty Propagation in the FIR Response}
\label{subsec:UP_FIR}
Equation~\ref{eq:FIR_Resp} defines the output of the FIR filter as the discrete convolution of the input signal and the filter coefficients.
It can also be restated as a weighted sum of the input sequence of uniformly distributed random variables  $X_i \sim \mathcal{U}\left(\mu_i-\frac{\delta}{2},\mu_i+\frac{\delta}{2}\right)$ as
\begin{equation}
Y_{n} = \sum_{i=0}^{N} b_i X_{n-i} .
\end{equation}
$Y_n$ is then a random variable with mean
\begin{equation}
\mu = \sum_{i=0}^{N} b_i \mu_{i-n}
\end{equation}
and variance
\begin{equation}
\sigma^2 = \sum_{i=0}^{N} \frac{b_i^2 \delta^2}{12} .
\end{equation}

We can then calculate the output uncertainty directly using the closed-form solutions derived by Kamgar-Parsi, Kamgar-Parsi, and Brosh in 1995~\cite{Kamgar-Parsi1995}, which we restate using our own notation below. 
First, we define a Heaviside step function as
\begin{equation}
\Theta \left(t\right) = 
	\begin{cases}
	0 & \text{if } t < 0 \\
	\frac{1}{2} & \text{if } t = 0 \\
	1 & \text{if } t > 0
	\end{cases} .
\end{equation}

The probability density function (pdf) of $Y_n$ was derived as
\begin{equation}
\label{eq:PDF}
\begin{split}
f_{Y_n}\left(y_n\right) = &\frac{\left(-1\right)^{N+1}}{N!\,\tilde{b}} \sum_{s_0 = \pm 1} \cdots \sum_{s_N = \pm 1} \tilde{s} \\
&\times \left(y_n - \mu - S\right)^N \times \Theta \left(y_n - \mu - S\right)
\end{split}
\end{equation}
where
\begin{equation}
\tilde{b} = \prod_{i=0}^{N}b_i \delta , 
\end{equation}
\begin{equation}
\tilde{s} = \prod_{i=0}^{N}s_i , 
\end{equation}
and
\begin{equation}
S = \sum_{i=0}^N \frac{s_i b_i \delta}{2} . 
\end{equation}

This direct calculation quickly becomes intractable for large-order filters, as the nested sums lead directly to combinatorial explosion. 
However, the original authors~\cite{Kamgar-Parsi1995} also demonstrated that the distribution can be estimated by calculating Equation~\ref{eq:PDF} using the $M < N$ most dominant terms.

\section{Simulations}
\label{sec:Simulation}
We now evaluate the uncertainty propagation in FIR filters using a practical example.
Filtering is a common pre-processing step in many signal processing applications to focus specifically on frequency bands of interest.
For example,  electroencephalography (EEG) data contains information from multiple frequency bands of brain signals.
It is often useful to examine information from only specific frequency bands, particularly when the data application is known to be associated with specific frequency bands.
For example,  motor imagery (MI) tasks are associated with Event-Related Desynchronization (ERD) patterns in both the mu and beta bands within the motor cortex~\cite{McFarland2000,terHorst2013,Yu2022}. 
This corresponds to the frequency range of approximately 7\,Hz to 35\,Hz so signal processing applications investigating MI tasks will often begin pre-processing with a bandpass filter.

\begin{figure}
\centering
\includegraphics[width=2.5in]{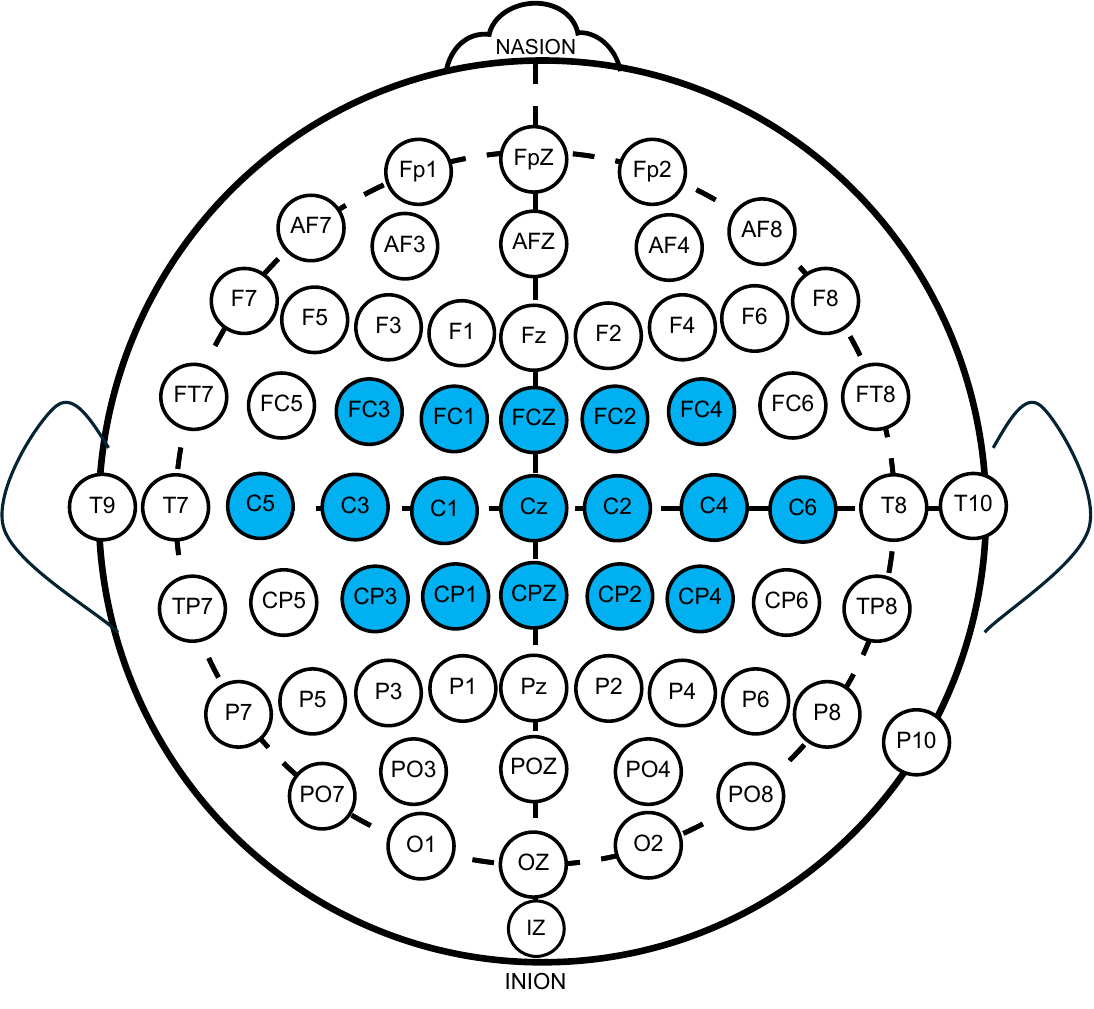}
\caption{Electrode placement in EEGBCI Data from Physionet~\cite{Schalk2004a} with motor imagery band channels highlighted in blue.}
\label{figure:electrode_placement}
\end{figure}

To demonstrate the effect of quantization uncertainty on the output of an FIR filter, we directly simulate the process using the EEGBCI dataset~\cite{Schalk2004a} pulled from Physionet~\cite{Goldberger2000}. 
In this dataset, subjects wore an EEG headset with 64 electrodes placed according to the configuration shown in Fig. \ref{figure:electrode_placement}. 
The subjects sat in front of a screen where a target appeared on either the right side or the left side of the screen and were instructed to imagine opening and closing their corresponding fist until the target disappeared. 
The EEGBCI dataset has a sampling frequency of 160\,Hz and an amplitude resolution of 1\,\textmu V.

We simulate the propagation of quantization error in EEG data by using the MNE~\cite{Gramfort2013} toolbox,  a commonly-utilized open-source EEG toolbox written in Python.
We first simulate quantization error of the raw EEG data by adding quantization noise pulled from a uniform distribution $\mathcal{U}(-0.5\,\text{\textmu V},0.5\,\text{\textmu V})$.
We then filter the data using the built-in FIR filter function in MNE. 
The filter used has a lower cutoff frequency of 7\,Hz with a 2\,Hz lower transition bandwidth and a higher cutoff frequency of 35\,Hz with a 8.75\,Hz transition bandwidth.
It uses a Hamming window to weight the coefficients.
The filter length is 265 samples\footnote{The filter MNE created is indeed 265 elements (not 256) because it is dependent on the transition bandwidths and sampling frequency. In particular it is equal to 3.3 times the longest transition period (in this case 0.5\,s) with the number of samples rounded up to the nearest odd integer}, which corresponds to 1.656 seconds.
We repeat this simulation 5000 times and obtain 561 filter output distributions per EEG channel per task epoch.

We compare the output of the filter to a scenario where no quantization uncertainty is added and refer to that as the post-filter error. 
This post-filter error should have the distribution described in Equation~\ref{eq:PDF} with $\mu = 0$.
If we attempt to use all 265 filter coefficients to calculate the distribution in Equation~\ref{eq:PDF}, the calculation would be intractable. 
However,  many of the filter coefficients have a magnitude close to 0, and the output of the filter is dominated by a relatively low number of coefficients.
As we previously noted in Section~\ref{subsec:UP_FIR} we can calculate the approximate distribution by only using the most dominant coefficients. 
We therefore calculate an approximation of the theoretical PDF using only the coefficients whose magnitudes are at least 5\% of the magnitude of the most dominant coefficient.
This reduces our number of coefficients to 19.

\begin{figure*}
\centering
\subfloat[]{\includegraphics[width=2in]{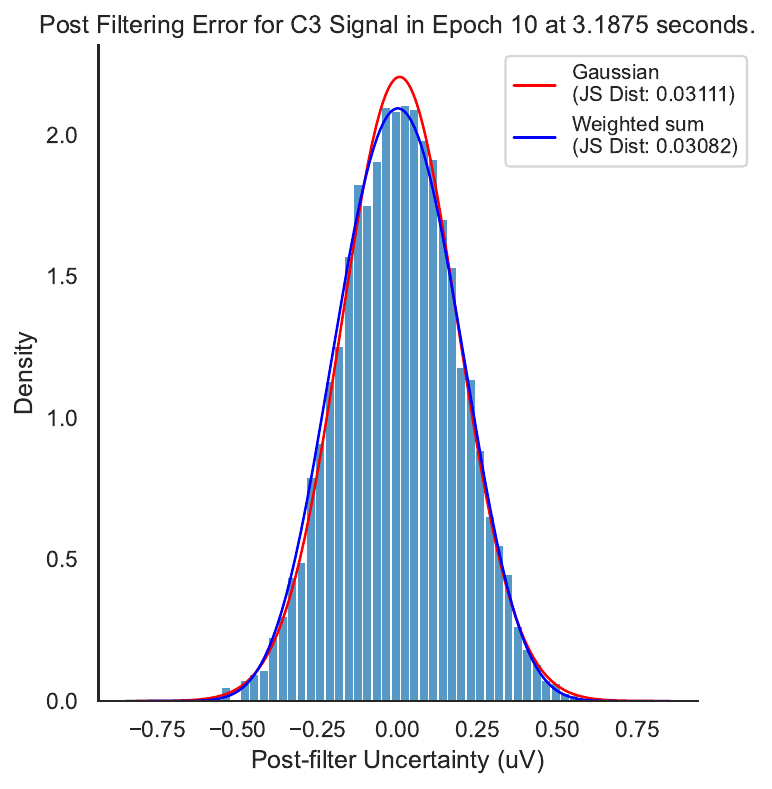}}
\hfil
\subfloat[]{\includegraphics[width=2in]{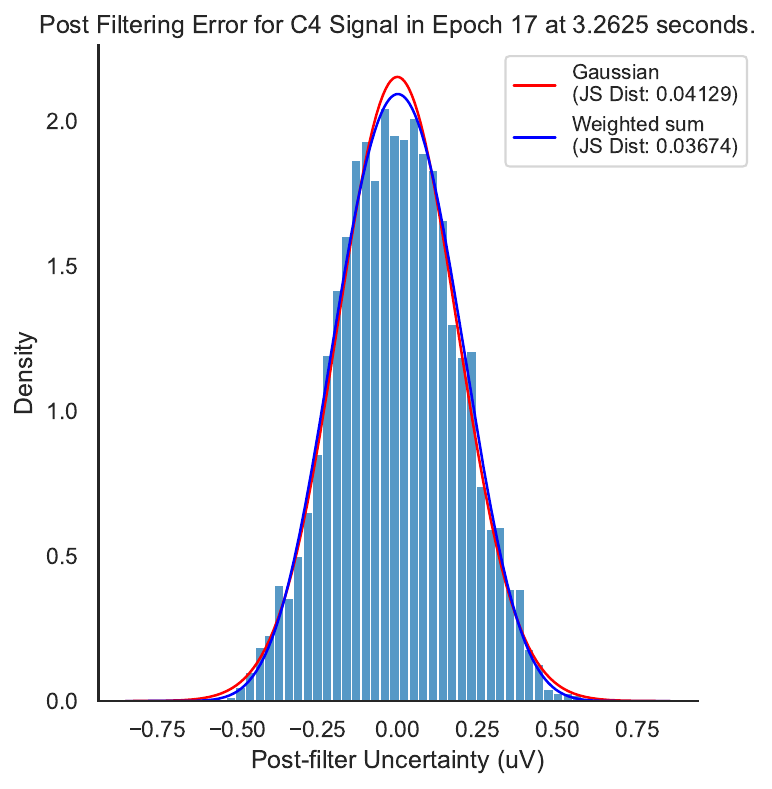}}
\hfil
\subfloat[]{\includegraphics[width=2in]{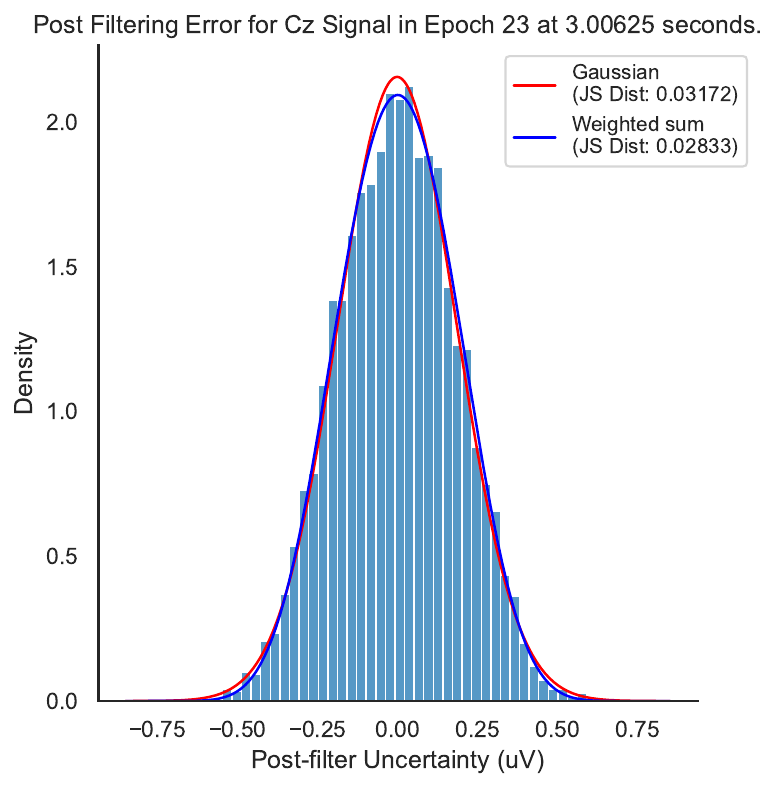}}
\vfil
\subfloat[]{\includegraphics[width=2in]{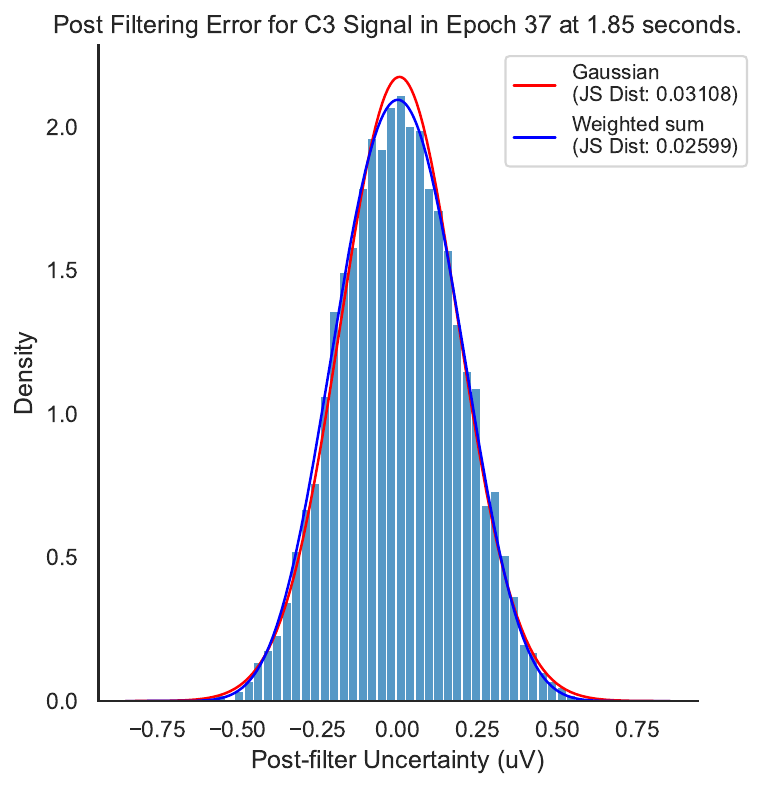}}
\hfil
\subfloat[]{\includegraphics[width=2in]{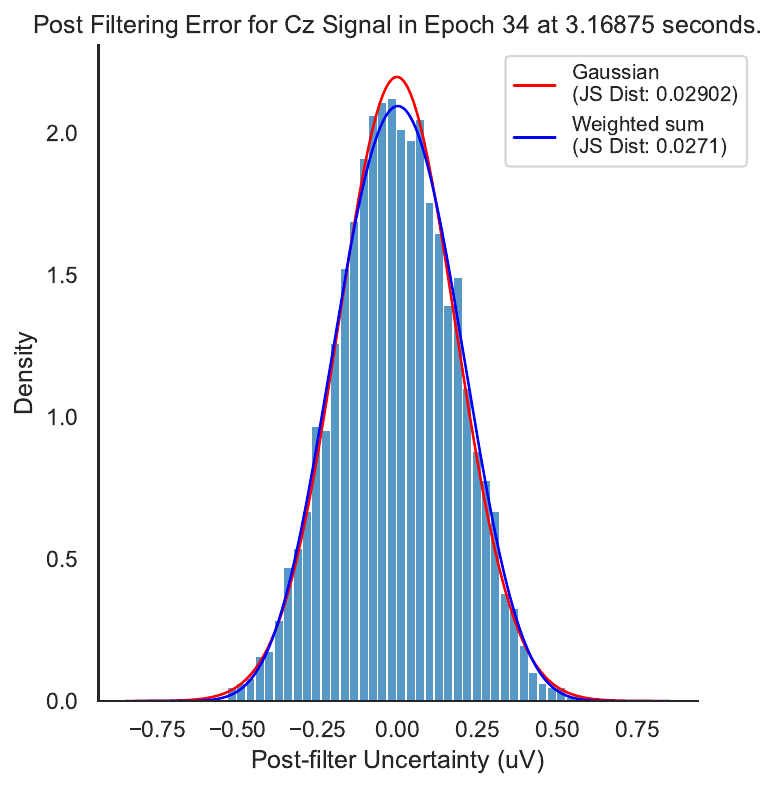}}
\hfil
\subfloat[]{\includegraphics[width=2in]{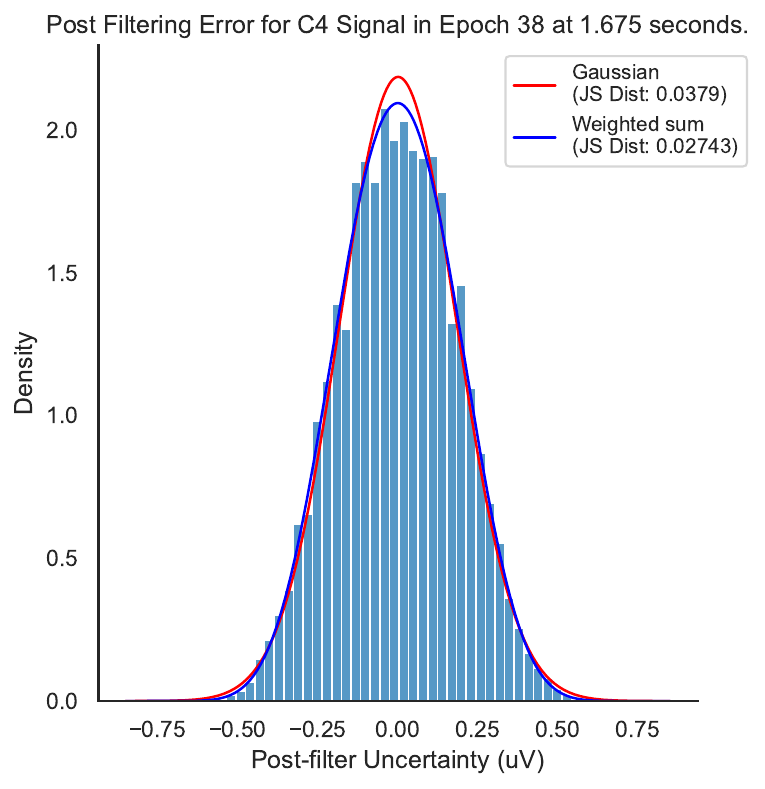}}
\vfil
\subfloat[]{\includegraphics[width=2in]{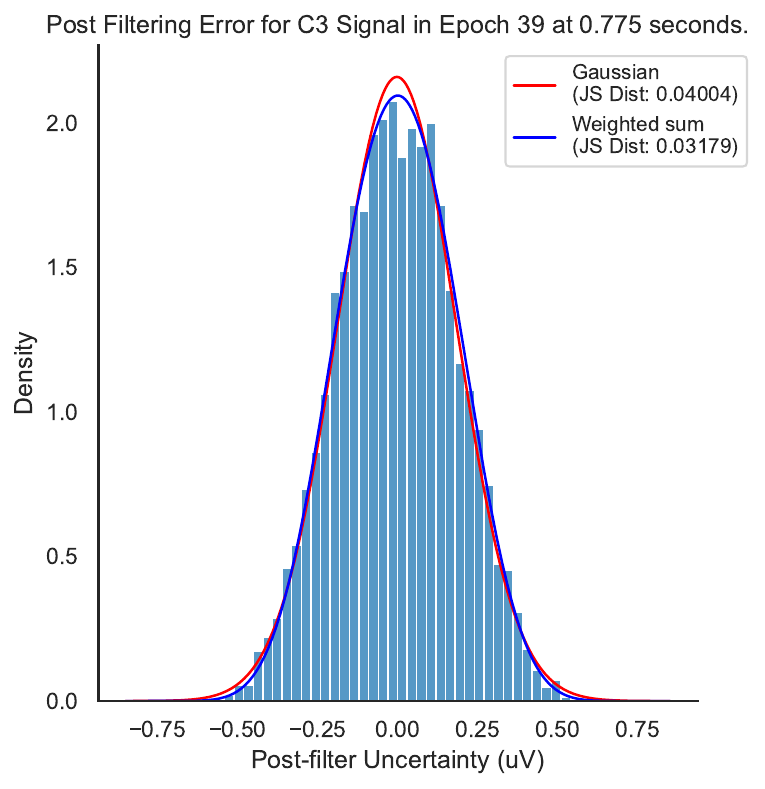}}
\hfil
\subfloat[]{\includegraphics[width=2in]{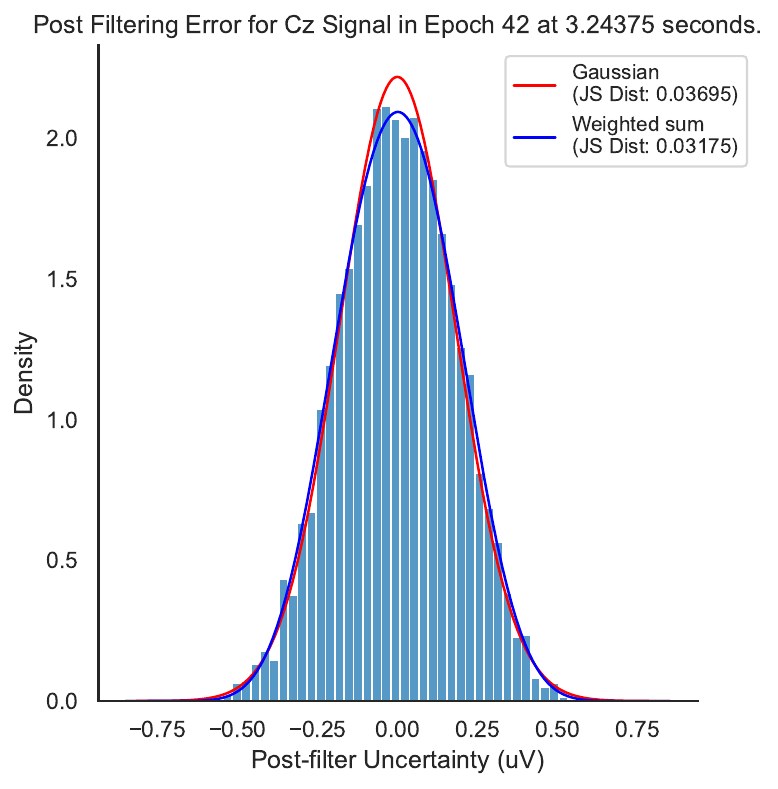}}
\hfil
\subfloat[]{\includegraphics[width=2in]{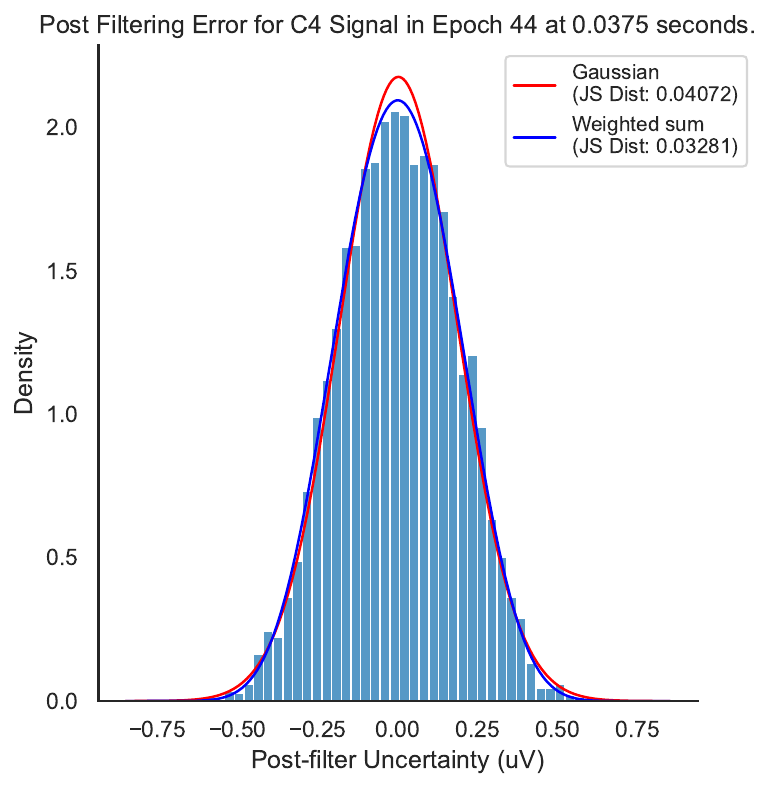}}
\caption{A selection of output uncertainty distributions across multiple epochs, channels, and timestamps.The bars represent the data distribution, the red line is the pdf of a Gaussian distribution with the same mean and variance, and the blue line is the pdf of the weighted-sum distribution calculated with only the most dominant coefficients from the FIR filter.From observation, our simulated data does not conform to the theoretical Gaussian distribution, but does line up well with the weighted-sum distribution calculated with just the most dominant coefficients}
\label{fig:SimResults}
\end{figure*}

We compare each of the simulated output distributions to both the theoretical pdf from Equation~\ref{eq:PDF} and to a theoretical Gaussian with the same mean and variance as the output.
Fig.~\ref{fig:SimResults} shows a small selection of output uncertainty distributions across multiple epochs, channels, and timestamps.
In total we have an output of 75,735 distinct output distributions.
We use the D'Agostino \& Pearson test~\cite{DAgostino1971, DAgostino1973} against the null hypothesis that the output distributions. 
The null hypothesis was rejected in 99.7\% of the output distributions with a p-value of 0.05 or less. 
The output distributions are all platykurtic with an average excess kurtosis value of $-0.2924$, furthering the evidence that the output distributions did not quantitatively conform to a Gaussian distribution.

We compare the Jensen-Shannon distances between the output uncertainty distributions and the two theoretical pdfs. 
The Jensen-Shannon distance between the output uncertainty distributions and the weighted sum pdf was consistently smaller than the Jensen-Shannon distance between the output uncertainty distributions and the theoretical Gaussian distributions.
In particular, the average distance between the output and the weighted sum distribution is $0.03184$ and the average distance between the output and the Gaussian distribution is $0.03661$.
We also note that while approximately $38.29\%$ of the values in Gaussian distribution lie within half of a standard deviation around the mean, approximately $36.81\%$ of the output distribution values lie within a half standard deviation around the data means.
Approximately $37.00\%$ of values of the weighted sum distribution lie within a half standard deviation around its mean.

We therefore conclude that the simulated output uncertainty distribution is better represented by the weighted sum pdf described in Equation~\ref{eq:PDF} than by the estimated Gaussian distribution.
However, the Jensen-Shannon distances and the percentage of the distribution near the mean are very similar for both the Gaussian distribution and the weighted sum distribution.\
Therefore, while the output uncertainty distribution of the filter is definitively non-Gaussian, a Gaussian estimate may still be useful when evaluating uncertainty propagation further along the signal processing chain.

\section{Conclusion}
\label{sec:Conclusion}
We analyzed the response of an FIR filter to inputs that contained measurement uncertainty due to quantization.
We expressed the output uncertainty of the FIR filter in terms of the input uncertainty and the filter's coefficients and showed that it cannot be assumed to be Gaussian.
We then ran simulations using open source EEG data to show that the output uncertainty of the FIR filter is generally sub-Gaussian and can be well approximated by using only the most dominant coefficients.

Even though our example application showed that the output uncertainty distribution of the FIR filter is non-Gaussian, the Jensen-Shannon distance between the simulated output distribution and the theoretical Gaussian distributions is small and comparable to the distance between the simulated output distribution and the theoretical weighted-sum distribution.
Therefore, depending on later processing steps, a Gaussian distribution may still be a useful model of the output uncertainty distribution. 
Researchers and practicing engineers should evaluate their specific applications to determine whether the difference between the true output distribution and the best-fitting Gaussian distribution are negligible before accepting the Gaussian assumption.

\bibliographystyle{IEEEtran}
\bibliography{IEEEabrv,references}

\end{document}